\documentclass[aps,twocolumn,floatfix,showpacs]{revtex4}
\usepackage{graphicx,bm}
\begin{document}
\title{Modeling interactions for resonant $p$-wave scattering}
\author{Ludovic Pricoupenko}
\affiliation
{Laboratoire de Physique Th\'{e}orique de la Mati\`{e}re Condens\'{e}e, 
Universit\'{e} Pierre et Marie Curie-Paris 6, 4 place Jussieu,
75252 Paris Cedex 05, France.}
\date{\today}
\begin{abstract}
In view  of recent experiments on  ultra-cold polarized  fermions, the
zero-range  potential approach  is   generalized  to situations  where
two-body scattering is resonant in  the $p$-wave channel. We introduce
a modified  scalar product which reveals a  deep relation  between the
geometry  of   the Hilbert space  and   the  interaction.  This
formulation is used to obtain a simple  interpretation for the transfer  rates
between atomic  and molecular states within  a two branches picture of
the many-body  system close to resonance.  At resonance, the energy of
the dilute gas is found to vary linearly with density.
\end{abstract}
\maketitle

Studies  of regimes with strong  correlations  is actually one of  the
most  challenging directions in   the field of   ultra-cold  atoms.  A
spectacular example   of such  a  situation  is  given by  the BCS-BEC
crossover of a  two spin-component Fermi gas,  obtained via tuning the
scattering length of binary collisions by using Feshbach resonances
\cite{unitary}. Another interesting example is resonant scattering in 
$p$-wave     channels  \cite{Regal,Zhang,Schunck,Ticknor,Chevy}   with
spin-polarized $^6$Li and $^{40}$K  atoms which opens the  possibility
of a BCS-BEC crossover in higher angular momentum channels in the near
future.  The major interest in these systems stems  from the fact that
the strength  of the correlations  can be  tuned arbitrarily while the
mean  inter-particle distance remains large with  respect to the range
of interatomic   forces.    Consequently, low energy   properties  are
independent  of the non-universal  short   range physics.   Ultra-cold
gases  can therefore  be  considered as   model  systems for  accurate
studies   of quantum  many-body  problems  underlying many interesting
phenomena  in condensed    matter physics.    For two   spin-component
fermions  where the   scattering amplitude  in  the $s$-wave   channel
dominates, several effective    two-body potentials are  used both  in
computational and analytical studies of the so-called unitary regime
\cite{Heiselberg,Combescot,Carlson,Astrakha,Yvanscaling,DePalo},  where
the scattering  length   diverges.  In  this   context, the zero-range
pseudo-potential  approach is very  appealing as  it  captures all the
effects of the interaction with only one parameter
\cite{YvanLesHouches,Vlambda,Yvanscaling} for broad  $s$-wave Feshbach
resonances,  and with   two parameters  in  narrow $s$-wave   Feshbach
resonances       \cite{Petrov3B}.     Furthermore, this   approach  is
particularly useful to obtain exact solutions of few body problems
\cite{Petrov3F,Petrov3B,Petrovscatt,Felix}, and also to improve 
unavoidable approximations in many-body systems~\cite{Vlambda,HFB2D}.

In this letter, a   general zero-range treatment of resonant  $p$-wave
interactions is introduced. This formalism  is used in the context  of
ultra-cold spin-polarized fermions,  where binary  $s$-wave scattering
is suppressed due  to the Pauli   principle. We show that  the Hilbert
space   associated with the zero-range  Hamiltonian  has to be defined
with a new metrics.  The modified scalar product illustrates nicely an
intrinsic connection between the geometry of the Hilbert space and the
interaction. The formalism is applied  to an effective model to obtain
some physical insight into the crossover  regime of a many-body system
made of particles  interacting in pairs  via the $p$-wave  channel.  A
two branches  picture is obtained   for the equation   of state in the
neighborhood of the Feshbach resonance, giving a simple interpretation
for   the transfer  rates  between atomic   and molecular states.   At
resonance, in the  dilute regime,  the  model predicts a  ground state
energy which varies linearly with respect to the atomic density.

Without  any  loss   of  generality,  we introduce   the  formalism by
considering two identical particles    in   absence of an     external
potential.      The      particles     of     mass    $m$,   positions
($\vec{r}_1,\vec{r}_2$)       and           relative       coordinates
${\vec{r}=\vec{r}_1\!-\!\vec{r}_2}$  are  described  in their center  of
mass   frame  by  the wave   function $\Psi(\vec{r}\,)$.  The interaction
between the particles  occurs only in the  $p$-wave channel of the
two-body  system and     is  modeled  by  the    following  zero-range
pseudo-potential:
\begin{equation}
\langle \vec{r}\, | V | \Psi \rangle 
= -  \frac{12 \pi \hbar^2 {\mathcal V}_s}{m} (\vec{\nabla} \delta)(\vec{r}\,)\,.\,\vec{\mathcal R}[\Psi\,]
\label{eq:V0} \quad ,
\end{equation}
where $\vec{\mathcal R}[\,.\,]$ is a regularizing operator defined by:
\begin{equation}
\vec{\mathcal R}[\Psi\,] = \lim_{r \to 0} \left[ (\frac{\partial_r^3}{2} 
+ \alpha \partial_r^2 ) r^2 \!\!\! \int_{\mathcal S_r}\!\! 
\frac{d^2\Omega\,\hat{e}_r}{4\pi} \Psi(\vec{r}\,)\right].
\label{eq:regularizing}
\end{equation}
In Eq.(\ref{eq:V0}), ${\mathcal V}_s$ is the scattering volume. In the
resonant  regime,  $|{\mathcal V}_s|$   is  arbitrarily large   and in
contrast with the  analog  unitary regime in  the  $s$-wave channel, a
second parameter (denoted here by $\alpha$) is essential for a description
of  the shape of  the two-body scattering  amplitude \cite{Landau} and 
has been introduced  in the expression  of  the regularizing operator.  
In Eq.(\ref{eq:regularizing}),  a surface  integration  is performed over
the sphere ${\mathcal S}_r$ of  radius $r$ centered at $r=0$, $d^2\Omega$
is  the elementary solid   angle and ${\hat{e}_r   =  \vec{r}/r}$.  This
integration ensures    that the  pseudo-potential  acts   only  on the
$p$-wave   component of  the   wave  function \cite{projection}. As  a
consequence,  the wave function  has a  $1/r^2$ singularity for  $r\to0$
which is a general feature of the zero-range potential approach in the
$p$-wave channel.  The role of the pseudo-potential in the Schr\"odinger
equation is two-fold: first,  it imposes a specific boundary condition
for the wave function in     the  vicinity of the $1/r^2$     $p$-wave
singularity:
\begin{equation}
\lim_{r\to  0} \left[ \left( {\mathcal V}_s \partial_r^3  + 2 \alpha {\mathcal V}_s
\partial_r^2 + 2 \right) r^2 \!\!\int_{\mathcal S_r}\!\!\!d^2\Omega\,\hat{e}_r \, \Psi \right] = \vec{0} \quad ,
\label{eq:Bethe_Peierls}
\end{equation}
and, second, it cancels  a diverging term proportional to $(\vec{\nabla}\delta)$
coming from the action of the Laplacian on the wave function. Equation
(\ref{eq:Bethe_Peierls})  is analogous    to the  contact    condition
introduced by H.~Bethe and   R.~Peierls for a description of  $s$-wave
scattering \cite{Bethe} and represents an alternative way to formulate
the zero-range approach.   The pseudo-potential in  Eq.(\ref{eq:V0})
generalizes    other      zero-range    $p$-wave     pseudo-potentials
\cite{Blume,Stock}, where the resonant regime must be described through
an energy dependent  scattering volume.  Contrary to these approaches,
the pseudo-potential   (\ref{eq:V0})  can  be used directly    in time
dependent  problems or in   situations where the two-body  collisional
energy is not explicitly defined.

As a  first  illustration of  the  formalism,  we deduce the  two-body
eigenstates of energy $E$ in their center  of mass frame in absence of
any  external  potential.  For  this  purpose, we  use   the  integral
equation:
\begin{equation}
\Psi(\vec{r}\,) = \Psi_0(\vec{r}\,) - \int\!\!d^3\vec{r}\,'\, 
G_E(\vec{r},\vec{r}\,') 
\langle \vec{r}\,'\, |V| \Psi \rangle \quad ,
\label{eq:green}
\end{equation}
where  $G_E$  is the   one-body  outgoing Green's  function at  energy
$E$.  For positive  energies, ${E=h^2k^2/m}$, Eq.(\ref{eq:green}) yields
the following expression for the scattering states:
\begin{equation}
\Psi_{\vec{k}}(\vec{r}\,) = \exp(i \vec{k}.\vec{r}\,) + 3 {\mathcal V}_s\, \hat{e}_r.\vec{\mathcal R}[\Psi_{\vec{k}}] \,
\partial_r\!\left(\frac{\exp(ikr)}{r}\right).
\label{eq:Green}
\end{equation}
Applying the regularizing operator ${\vec{\mathcal R}[\,.\,]}$ on both
sides  of Eq.(\ref{eq:Green}) solves the problem with:
\begin{equation}
\vec{\mathcal R}[\Psi_{\vec{k}}] = \frac{i \vec{k}}
{1+\alpha k^2{\mathcal V}_s+ik^3{\mathcal V}_s} \quad .
\label{eq:dipole}
\end{equation}
In the  asymptotic limit ($kr \gg 1$), one obtains:
\begin{eqnarray}
&&\displaystyle \Psi_{\vec{k}}(\vec{r}\,) \simeq \exp(i \vec{k}.\vec{r}\,) + 3\, (\hat{e}_k.\hat{e}_r) \, f_1 \, \frac{\exp(ik r)}{r} ,
\label{eq:scatt_states}\\ 
&&\displaystyle \mbox{with , } - \frac{1}{f_1} =  \frac{1}{k^2 {\mathcal V}_s} + \alpha  + i k ,\ \ \mbox{and}\  \hat{e}_k=\vec{k}/k . 
\label{eq:scatt_amplitude}
\end{eqnarray}
Eq.(\ref{eq:scatt_amplitude})   coincides  exactly  with   the general
expansion of the inverse $p$-wave scattering amplitude $f_1$ at second
order  in  the  low   energy limit   \cite{Landau}, showing   that the
pseudo-potential (\ref{eq:V0})  provides  a modeling  of two-body
collisions  for  relative  momenta   ${k\ll\alpha}$. The  zero-range  potential
approach requires also that ${k R \ll 1}$  where $R$ denotes the potential
range, that  is the radius  such that for  $r>R$ the  'real' potential
term experienced by particles can be  neglected in the Schr\"{o}dinger
equation.  In actual experiments on spin-polarized fermions
\cite{Regal,Zhang,Schunck} an   external magnetic field  $B$ tunes the
energy of a two-body bound state in a closed  channel which is coupled
to the  open  channel associated  with   the two  colliding  atoms.  A
Feshbach resonance occurs  for a vanishing   value of the  bound state
energy at  a  magnetic field  $B=B_0$.   Close to  the  resonance, the
scattering volume takes  arbitrarily large values with ${{\mathcal V}_s
\propto - 1/(B-B_0)}$,  while the parameter $\alpha$  is a slowly varying function
of $B$  \cite{Chevy,anisotropy}.   The  two-body  bound state   in the
closed channel extends on short lengths of the  order of the potential
range and is  not described by the  zero-range approach.  However,  in
the regime  where  ${\alpha^3 {\mathcal V}_s}$   is large and  positive ($B \lesssim
B_0$), the  pseudo-potential supports a  weakly  bound state of energy
${\epsilon_B \simeq -\hbar^2/m\alpha{\mathcal V}_s\propto  B-B_0}$ which corresponds to the outer
part  (region $r  >  R$) of the   molecular  state resulting from  the
coupling between the closed and the open channels.  We anticipate that
this state is populated by pairs of particles in the BEC region of the
BEC-BCS crossover  regime.  For large   and  negative values  of  ${\alpha^3
{\mathcal V}_s}$  (${B \gtrsim B_0}$), this state transforms  into a long lived
quasi-bound state \cite{Landau}. 

Having  in hand  the short  range behavior of  the   wave functions in
Eq.(\ref{eq:Green}), one may wonder  what is the underlying  structure
of  the Hilbert   space spanned by  the  two-body  eigenstates of  the
zero-range  Hamiltonian.  As a   consequence  of the ${1/r^2}$ singularity, 
the low energy bound state is not normalizable. Moreover,
the usual scalar product between two different scattering states gives
an infinite result.  Nevertheless, as we  show in the following, it is
possible to introduce  a  regularized  scalar  product such  that  the
eigenfunctions are orthogonal to each others and  the low energy bound
state is normalizable.   For   this purpose,  we perform    the scalar
product between two states ${|\Psi_{\vec{k}}\rangle}$ and ${|\Psi_{\vec{k}\,'}\rangle}$ with
${\vec{k} \neq \vec{k}\,'}$, but exclude  from integration the inner volume
of the sphere ${\mathcal S}_r$, defined above. One finds, as ${r \to 0}$:
\begin{eqnarray}
\int_{r'>r}&&\!\!\!\!\!\!\!d^3\vec{r}\,'\, \Psi_{\vec{k}}^*(\vec{r}\,') 
\Psi_{\vec{k}'}(\vec{r}\,') \nonumber \\ 
=&&\!\!\! (r-\alpha r^2) \! 
\int_{{\mathcal S}_r} \!\!\! d^2\vec{r}\,'\, 
\Psi_{\vec{k}}^*(\vec{r}\,') \Psi_{\vec{k}'}(\vec{r}\,') 
+ {\mathcal O}(r) .
\label{eq:reste}
\end{eqnarray}
This integral is non zero and diverges in  the zero-range limit due to
the  $p$-wave singularity  of  the  scattering states.   Therefore the
zero-range   Hamiltonian is not hermitian  with  respect  to the usual
scalar product.   This feature follows from the  fact that the mapping
between the \emph{true} scattering  states associated with  the finite
range  potential    experienced    by  particles  and    the    states
$\left\{|\Psi_{\vec{k}}\rangle\right\}$ in Eq.(\ref{eq:Green})  is not  valid for
$r \lesssim R$ while it is justified outside the potential range. Indeed, the
singular     boundary  behavior   imposed     on   wave  functions  in
Eq.(\ref{eq:Bethe_Peierls}) is the way to reproduce the effect of the
\emph{true} finite  range  potential  for  $r>R$  but has    a  formal
character for interparticle distances  $r \lesssim R$. In order to conciliate
the zero-range pseudo-potential with a finite range potential of small
radius $R$, we consider  the sphere $S_R$   which separates the  outer
part (region $r >  R$)  and the inner part   (region $r \lesssim R$)  of  the
\emph{true} wave functions.  Contribution in the scalar product of the
outer parts  of the     \emph{true}    wave functions  is    given   by
Eq.(\ref{eq:reste})  with $r=R$ and  due to  orthogonality between the
\emph{true} scattering states, the scalar  product of the  inner parts
cancels  this   term. By construction,   the inner  parts  of the wave
functions   ($r \lesssim R$) are   not    described in  the  zero-range
approach. However, we are free  to modify the  usual scalar product in
order to take into account their contribution.  Therefore, we define a
regularized   scalar   product $(\,.\,|\,.\,)_0$   by subtracting  the
surface term  (\emph{r.h.s.}  of  Eq.(\ref{eq:reste})) from  the usual
scalar product.  In the    formal zero-range limit, the    regularized
scalar product can be also written with a weight $g(r)$~given~by:
\begin{equation}
g(r) = 1 + \delta(r) \left[(\alpha r^2-r)\,.\,\right] \quad ,
\label{eq:weight}
\end{equation}
so  that finally, for wave   functions obeying the boundary  condition
Eq.(\ref{eq:Bethe_Peierls}), the scalar product reads:
\begin{equation}
(\Psi|\Psi')_0 = \int\! d^3 \vec{r}\,\, g(r) \Psi^*(\vec{r}\,) 
\Psi'(\vec{r}\,) \quad .
\label{eq:regularized_scalar_product}
\end{equation}
One  can check that with  this new metrics, the zero-range Hamiltonian
is  hermitian  in the    domain defined by   wave-functions satisfying
Eq.(\ref{eq:Bethe_Peierls}).  Note that  following the same reasoning,
the notion of regularized scalar product has been generalized recently
for zero-range interactions in all the other partial wave channels
\cite{lwaves} in the resonant regime.  Coming  back to the low energy   bound state of energy
${\epsilon_B= -\hbar^2\kappa_B^2/m}$ with ${\kappa_B^{-2} \simeq \alpha {\mathcal V}_s}$ appearing in the
regime  ${\mathcal V}_s \alpha^3\gg  1$,  the   radial part normalized   with
respect to Eq.(\ref{eq:regularized_scalar_product}) is:
\begin{equation}
{\mathcal R}_B(r) = \frac{1}{\sqrt{\alpha - 3\kappa_B/2}}\,\, \partial_r\!\left(\frac{\exp(-\kappa_B r)}{r}\right) \quad .
\label{eq:bound_state}
\end{equation}
As  a result,  the probability  of  finding the molecule  of vanishing
binding energy outside the range of the true potential is of the order
of $1/\alpha R$, meaning that the approach is consistent for $\alpha>0$ and $\alpha R
\gtrsim 1$ (for example $\alpha R \simeq 2.8$ in  $^{40}$K for the two resonances with
$B_0 \simeq  198.5$~G   \cite{Ticknor}).   The  normalization   factor   in
Eq.(\ref{eq:bound_state}) can be also deduced  from the residue of the
scattering amplitude (\ref{eq:scatt_amplitude})   at  the energy  ${\hbar^2
k^2/m=\epsilon_B}$ \cite{Landau} and the  regularized scalar product extends this method to inhomogeneous situations.

Now  we   turn to the case   of   a homogeneous  system   made  of $N$
spin-polarized identical fermions  of mass $m$.   For  a density $(n)$
sufficiently small that binary processes  occur at low energy  (${nR^3\ll
1}$) the  zero-range approach is justified  and the interaction between
two given  particles can   be  modeled  by  the  pseudo-potential in
Eq.(\ref{eq:V0}) \cite{lambda}. For  a qualitative  picture  of the  
physics involved in  the
neighborhood of  the resonant regime,  we use the box model introduced
in Ref.\cite{ToyBox}  in the context of  $s$-wave scattering.  In this
effective approach, the interaction of one fermion with all the others
is modeled by  the interaction of a fictitious  particle of mass equal
to the reduced mass $m/2$, with  a fixed scatterer  at the center of a
spherical  box  of radius $L$.   The wave   function of the fictitious
particle which is   non-zero   in the  $p$-wave  channel  only  is  an
eigenstate of the pseudo-potential  (\ref{eq:V0}) and vanishes on
the surface   of the box.   It  represents the  pair  function  of two
fermions in their center  of mass  frame.   The boundary condition  at
$r=L$, mimics the effects of correlations between pairs and the radius
$L$   is fixed  by considering  the  non  interacting  case.   In this
situation, the  total energy of  the  gas is ${E  = \frac{3}{5} N \epsilon_F}$,
where the Fermi energy ${\epsilon_F= \hbar^{2} k_F^{2}/2m}$ is  related to the mean
density $n$, through ${k_F^3=6\pi^2n}$. The gas energy $E$ is also related
to the energy $\epsilon$ of the fictitious particle by ${E=N\epsilon/2}$.  Finally, we
obtain a relation between the radius  $L$ and the Fermi momentum $k_F$
by considering the  ground state energy  of the fictitious particle in
the box where ${k_F L \simeq 5.8}$. Hence, $L$ is as expected of the order of
the  mean  inter-particle distance.   In   the interacting  case,  the
equation  of  state  is deduced   from  the energy  of  the fictitious
particle  ${\epsilon=\hbar^2k^2/m=2E/N}$, where $k$   (real for $\epsilon>0$ or  imaginary
otherwise) is a solution of the equation:
\begin{equation}
\frac{kL \cos(kL)-\sin(kL)}{kL\sin(kL)+\cos(kL)}= - \frac{{\mathcal V}_s k^3}{1+\alpha {\mathcal V}_s k^2} \quad .
\label{eq:eos}
\end{equation}
In current experiments \cite{Regal,Zhang,Schunck}, if we neglect  the trap
geometry,  the density and the external magnetic field are the two control
parameters.  We consider then solutions of Eq.(\ref{eq:eos}) for given
values  of ${-1/ {\mathcal V}_s \propto B-B_0}$ and of the  dimensionless parameter 
${\alpha/k_F \gg  1}$.
\begin{figure}[t]
\resizebox{!}{8cm}{\includegraphics{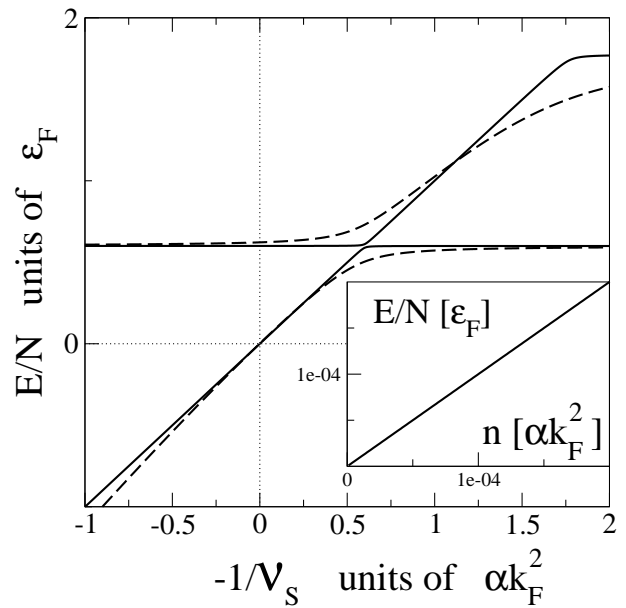}}
\caption{Energy per particle for a homogeneous medium composed of 
polarized fermions as a function of ${-1/{\mathcal V}_s}$ for ${\alpha=10 k_F}$ 
(dashed line) and ${\alpha=10^3\,k_F}$ (continuous line). Inset: equation of 
state at resonance ($B=B_0$). In this regime, the energy per 
particle varies linearly with the atomic density.}
\label{fig:ener}
\end{figure}
In Fig.(\ref{fig:ener}), we have plotted the two first branches of the
energy per particle as a function of $-1/{\mathcal V}_s$ for two values of 
the density (${\alpha/k_F=10}$  and ${10^3}$). The  left part of the  
upper branch corresponds to  the metastable  weakly   repulsive atomic 
phase.   The right   part  of the  ground  branch   is associated  
with the  weakly attractive atomic phase where a BCS  phase is expected 
at sufficiently low temperature.  The  left part of the  ground  branch represents the
molecular  phase  composed  of  dimers   of energy  
${\epsilon/\epsilon_F \simeq -2/{\mathcal V}_s \alpha k_F^2}$  \cite{anisotropy2}. As
explained in Ref.\cite{ToyBox}, this two branches picture provides the two
possible  scenarios for  obtaining a  molecular phase  from the weakly
interacting atomic  Fermi gas by  varying the external magnetic field.
In  the first  scenario the  system is initially  a  weakly attractive
Fermi gas and  follows the  ground  branch while the parameter 
$-1/{\mathcal V}_s$ is
tuned  from positive to negative values.   In the second scenario, the
system is initially a weakly repulsive Fermi gas  and is driven in the
resonant  regime by  increasing values  of  $-1/{\mathcal V}_s$. The 
ground branch is then   populated  by  three   body  recombination  
processes. At resonance ($B=B_0$)  the  ground
state energy  varies    linearly  with   density: 
${E/N \propto \hbar^2 n/m\alpha\ll \epsilon_F}$ 
(see inset of Fig.(\ref{fig:ener})).  This result  which has been found
also  in  a   BCS  treatment \cite{Jason}  strongly differs from  the 
$n^{2/3}$ law obtained in the unitary        regime  for two-spin  
components      fermions \cite{ToyBox}. Qualitatively,  
a  pair  of  interacting fermions which would have zero energy and 
infinite size at resonance in absence  of other particles, is
confined by  its neighbors in a  volume  of the order  of $L^3 \propto
1/n$ and the box model provides an estimation of the cost in energy of
this    configuration  which  is     ${\simeq  \hbar^2/m\alpha   L^3\ll
\hbar^2/mL^2}$.  Also  the  level  crossing between  the  two  branches
differs from  the $s$-wave result:  it is not  only  translated to the
right of the zero energy  resonance  (${B=B_0}$) but also its  amplitude
tends  to  zero for decreasing   densities.  As a consequence  one may
expect  that  in  the  region  of the  level  crossing, non  adiabatic
transfers of  atoms   from the atomic branch   to   the ground  branch
resulting  from inelastic three-body  collisions is followed by only a
small heating  of  the gas  due  to the  small  kinetic energy of  the
outgoing resulting states.

To conclude, we comment on the stability  of the molecular phase which
is one of the main issue for achieving the BEC-BCS crossover. Lifetime
of the shallow diatomic molecules is limited by recombinations into deep
bound  states  of energies  ${{\mathcal  O}(-\hbar^2/mR^2)}$. Following, the
analysis of Ref.\cite{Petrovscatt} the corresponding  loss rate can be
estimated from the probability that  three atoms are confined within a
volume of characteristic length $R$. The fact that the probability for
a pair  to be in  the inner part of the  potential ($r<R$) is given by
${\sim(1-1/\alpha R)}$ is  a strong  indication that this  rate depends 
on the width of the resonance considered and could be reduced in the 
cases where ${\alpha R \sim 1}$. In contrast, these losses are 
universally suppressed in the unitary two-spin components Fermi 
gas~\cite{Petrovscatt}.

{\bf Acknowledgments :} Y. Castin, F. Chevy, M. Holzmann and F. Werner
are acknowledged for thorough discussions on the subject. LPTMC is UMR
7600 of CNRS.


\begin{thebibliography}{99}
\bibitem{projection} This can be seen from the expression
of the projection operator in the $p$-wave space:
$3\hat{e}_r/4\pi \,. \int \! d^2\Omega \, (\hat{e}_r \,\Psi)$.
\bibitem{anisotropy} For sake of simplicity, we have neglected the small 
anisotropy in the  two-body interaction \cite{Ticknor}.   This feature
can be  implemented  in Eq.(\ref{eq:V0})  by the introduction of a set of
different  parameters $\{{\mathcal V}_s^i,\alpha^i \}_{i=x,y,z}$ associated
with the three components  of the vector ${\mathcal V}_s \vec{\mathcal
R}[\,.\,]$.
\bibitem{lambda} In view of applications of the formalism to approximate schemes
of the many-body   problem, the pseudo-potential (\ref{eq:V0}) can  be
generalized by a  family of equivalent potentials  generated by a free
parameter $\lambda$,  analogous  to  the   $\lambda$-potential introduced  in  the
$s$-wave channel \cite{Vlambda}: ${\displaystyle \langle \vec{r}\, | V_\lambda | \Psi
\rangle = -  \frac{12 \pi \hbar^2 {\mathcal V}_s}{m(1-\lambda {\mathcal V}_s)}  (\vec{\nabla}
\delta)(\vec{r}\,)\,.\,\vec{\mathcal R}_\lambda[\Psi\,]}$, where ${\displaystyle
\vec{{\mathcal R}}_\lambda[\Psi] = \vec{{\mathcal R}}[\Psi] + \frac{\lambda}{4\pi}
\lim_{r \to 0} r^2 \! \int_{\mathcal S_r}\!\!\! d^2\Omega\,\hat{e}_r \Psi(\vec{r}\,)}$.
While  exact  solutions can  be obtained  for any  value  of $\lambda$, some
approximations  are $\lambda$-dependent and   can be improved by choosing  a
particular value of the free  parameter \cite{Vlambda,HFB2D}. For instance
at  the  two-body  level, the exact  two-body  scattering  states  are
obtained  in the first  order  Born  approximation for  ${\lambda=(ik)^2(\alpha+i
k)}$. In many-body systems,  the dependence in the $\lambda$-parameter
which appears  in  the  Hartree-Fock-Bogoliubov approximation for  the
Bose gas \cite{Vlambda,HFB2D} is also present in the BCS approximation
for spin-polarized fermions.
\bibitem{anisotropy2} Due to the small anisotropy in the two-body interaction, 
the Feshbach resonance is split in a doublet \cite{Ticknor}, so that molecules 
in the BEC phase have a prefered orientation. For a quantification axis along 
the magnetic field $\vec{B}$, one can show from Ref.\cite{Ticknor} that the angular wave function 
of a dimer in the  ground branch of lowest energy is $Y_{10}(\Omega)$.
\end{thebibliography}
\end{document}